# Decoupling Crossover in Asymmetric Broadside Coupled Split Ring Resonators at Terahertz Frequencies

G. R. Keiser[1], A. C. Strikwerda[1,3], K. Fan[2], V. Young[1], X. Zhang[2], and R. D. Averitt[1]

[1]Boston University, Department of Physics, Boston, MA 02215, USA
[2]Boston University, Department of Mechanical Engineering, Boston, MA 02215, USA
[3]Technical University of Denmark, DTU Fotonik – Department of Photonics Engineering, Kgs. Lyngby, DK-2800, Denmark

**Abstract:** We investigate the electromagnetic response of asymmetric broadside coupled split ring resonators (ABC-SRRs) as a function of the relative in-plane displacement between the two component SRRs. The asymmetry is defined as the difference in the capacitive gap widths ($\Delta g$) between the two resonators comprising a coupled unit. We characterize the response of ABC-SRRs both numerically and experimentally via terahertz time-domain spectroscopy. As with symmetric BC-SRRs ($\Delta g=0\mu m$), a large redshift in the LC resonance is observed with increasing displacement, resulting from changes in the capacitive and inductive coupling. However, for ABC-SRRs, in-plane shifting between the two resonators by more than $0.375L_o$ ($L_o$=SRR sidelength) results in a transition to a response with two resonant modes, associated with decoupling in the ABC-SRRs. For increasing $\Delta g$, the decoupling transition begins at the same relative shift ($0.375L_o$), though with an increase in the oscillator strength of the new mode. This strongly contrasts with symmetric BC-SRRs which present only one resonance for shifts up to $0.75L_o$. Since all BC-SRRs are effectively asymmetric when placed on a substrate, an understanding of ABC-SRR behavior is essential for a complete understanding of BC-SRR based metamaterials.

Metamaterials present a vast array of exciting possibilities for optical materials engineering, allowing for the design and fabrication of materials where the electric permittivity, magnetic permeability, and impedance Z can be specified with ever-increasing precision. Split-ring resonators(SRRs), first proposed by John Pendry [1], are ubiquitous in metamaterial designs, appearing in negative index materials [2, 3], electromagnetic cloaks [4], memory metamaterials [5], thermal detectors [6], and perfect absorbers [7, 8] to name just a few representative examples. As the field has progressed, geometrical variants on the basic SRR have appeared in the literature [9-11]. The push for tunable and controllable metamaterial devices [12] has ignited interest in SRR variants where manipulation of near field coupling can be used to alter a MM response [13]. Manipulation of coupled MM structures has born much fruit in recent years, especially in the areas of tunable [14] and nonlinear MMs [15]. In particular, broadside coupled SRR's (BC-SRRs) [16] have attracted considerable attention due to their high structural tunability [13, 17] and ability to eliminate parasitic bianisotropic effects at the unit cell level [18].

To date, the majority of research on the transmission characteristics and tunability of BC-SRRs has focused on designs where both rings have identical resonance frequencies [17, 19]. Only one exception, a small subsection of [20], is known to the authors. There are, however, several ways in which broadside coupled resonators become effectively asymmetric, meaning that the two resonators have different resonant frequencies. For example, the presence of a substrate will break symmetry and induce bianistropy into any metamaterial design [21] including planar BC-SRRs. It is also possible to create asymmetric BC-SRRs (ABC-SRRs) by varying the relative geometrical parameters (and hence capacitance and inductance) of the SRRs that comprise the BC-SRR unit. Given the many ways in which asymmetry can be introduced into coupled resonators (either on purpose, or spuriously) systematic research into the electromagnetic properties of ABC-SRRs is vital to gain a more complete understanding of metamaterials.

We expand on the results of previous work on symmetric BC-SRRs [17, 19, 22] by investigating how an in-plane, lateral shift (see Fig. 1) between the two elements comprising an ABC-SRR affects the response of the metamaterial as a whole. This is accomplished by fabricating stand-alone ABC-SRR structures in polyimide where the asymmetry is defined by the difference in the gaps widths ($\Delta g$) in the two resonators. For a given $\Delta g$, the lateral shift is varied and the resulting electromagnetic response experimentally measured using terahertz time-domain spectroscopy (THz-TDS). Our electrically active metamaterials (i.e. resonant under excitation by the THz electric field) exhibit a large redshift in the fundamental mode as a function of increasing lateral shift between the SRRs. This effect is due to the shift induced change in near-field coupling between the SRRs comprising the BC-SRRs [17, 23]. Of particular significance, a second resonance appears for shifts greater than $0.375L_o$ where $L_o$ is the side length of the component SRRs (Fig. 1). This dramatic change of behavior arises as the ABC-SRR transitions to a "decoupled state" for large shift values, defined as a state where the component SRRs respond to incident radiation as *separate, individual resonators*. Full-wave electromagnetic simulations confirm this explanation. This behavior is in stark contrast to the behavior of a symmetric BC-SRR, which shows only one coupled mode for all shift values up to $0.75L_o$ [17]. We conclude by providing intuitive explanations for this disparate behavior, consistent with the conceptual models published in previous work.

The design of the ABC-SRR structures is outlined in Fig. 1. The unit cell of the metamaterial is composed of two square SRRs separated by a 5µm polyimide substrate ($\varepsilon_r$ = 2.88 loss-tangent $\tan(\delta)$ = 0.0313). The rings are rotated 180° relative to each other, producing a "broadside coupled" configuration (Fig. 1). Both rings are then covered with a 5µm polyimide superstrate. The dimensions of an individual ring are shown in Fig. 1(c). The unit-cell periodicity is $P$=60µm, metallization side-length $L_o$=40µm, metallization width $w$=11µm, front gap width $g$=2µm, and the back gap width varies from 4µm to 16µm in 4µm steps. The lateral shift ($L_{shift}$) between the two rings (Fig. 1(d)) varies from 0µm to 25µm in 5µm steps. The dimensions are such that a 30µm shift is equal to a shift of half a unit cell. These structures were fabricated using conventional photolithography as described in detail in [17].

Following fabrication, the metamaterials were characterized using terahertz time-domain spectroscopy (THz-TDS). The radiation was normally incident and oriented such that the electric field pointed perpendicular to the SRR capacitive gaps. This ensures that we are exciting the structures via the electric field and not the magnetic field since the incident magnetic field does not thread the SRRs (Fig. 1(d)).

The transmission as a function of frequency for structures with 0 µm lateral shift and 25 µm lateral shift are shown in Fig. 2(a) and (b), respectively. For each lateral shift, the data is shown for $\Delta g$ = 2, 6, 10, and 14µm. For the unshifted case (Fig. 2(a)) a single resonance is observed, corresponding to the expected electrical resonance [17]. However, in the 25µm shifted case, two modes are excited. A strong electrical resonance now appears at lower frequencies, and a second resonance appears at a higher frequencies. The strength of this second resonance is directly proportional to $\Delta g$. One possibility is that this second mode is the ABC-SRR's other, magnetic mode, excited through the bianisotropy inherent in the asymmetric resonator design. However, this is not the case. If it were, one would expect a resonance of approximately equal strength appear at lower frequencies for shifts smaller than $L_{shift}/L_o=0.375$ as well [19]. In the following, it is demonstrated that the appearance of the second mode arises as the individual SRRs comprising the ABC-SRR unit cell start responding to incident radiation separately, as individual, *uncoupled* resonators. Summarizing the results of Fig. 2, for unshifted resonators (Fig. 2(a)), varying $\Delta g$ causes no significant effects on the electromagnetic response. In contrast, for shifted resonators (Fig. 2(b), $L_{shift}>0.375L_o$), a new mode appears for $\Delta g > 0\mu m$, with a magnitude that increases with the magnitude of $\Delta g$.

In order to shed light on the phenomena involved, the electromagnetic response of the ABC-SRRs was modeled using CST Microwave Studio's frequency solver. The simulated transmission vs. $L_{shift}$ for multiple values of $\Delta g$ are presented and compared with experiment in Figure 3. As the figure demonstrates, two resonances are excited for shift values greater than $0.375L_o$, and the strength of this resonance increases as $\Delta g$ is increased.

Consider the trend of the single mode for $L_{shift}<0.375L_o$. This primary mode of the structure moves to lower frequencies as $L_{shift}$ is increased. This is due to the shift-induced change in mutual capacitance and inductance between the two SRRs as described in detail in [17] for the case of equivalent resonators ($\Delta g=0\mu m$). For $L_{shift}=0\mu m$, the mode is blueshifted from the bare resonance of a lone SRR due to capacitive and inductive coupling between the rings. For the resonance under consideration, the mutual inductance starts out negative and increases with shift. The mutual capacitance also increases with shift, since the positive charge distribution of one SRR is moved closer to the negative charge distribution on the other SRR. Thus, as the SRRs are shifted laterally, the total capacitance and inductance will increase, decreasing the resonance frequency since $\omega_o \sim 1/\sqrt{(LC)}$. The resulting effect on this mode is to redshift until it undergoes an avoided crossing with the magnetic resonance of the structure (not excited in the experimental configuration at normal incidence) at $0.375L_o$.

For further shifts, two electric modes are now present. The lower frequency resonance redshifts until $L_{shift}=0.75L_o$. The higher frequency resonance experiences a slight blueshift over the same range. For shifts greater than $0.75L_o$, the SRRs begin to overlap with the SRRs from neighboring unit cells, resulting in the reverse process with a corresponding blueshift for the low frequency mode and a corresponding redshift for the high frequency mode back to the resonance positions for $L_{shift}=0.375L_o$. For even larger shifts, the ABC-SRR transitions back to single mode behavior, with the mode blueshifting back to the resonance position at $L_{shift}/L_o=0$.

The appearance of the second high frequency mode for $L_{shift}/L_o>0.375$ is surprising. The mode is present only for large shift values, and has an oscillator strength strongly dependent on the asymmetry between the SRRs (Figs. 2(b) and 3), signaling the onset of new behavior not observed in previous work focusing solely on symmetric BC-SRRs.

Further insight into the physical mechanism behind the appearance of this higher frequency resonance becomes apparent by considering the response of an ABC-SRR structure with a larger unit cell. In this case, we have doubled the size in the y direction to create a rectangular unit cell with $P_y=2P_x=120\mu m$ while leaving all other SRR dimensions the same. This change will increase the space between each ABC-SRR and its nearest neighbor in the direction of $L_{shift}$. Figure 4(a) shows the simulated transmission response of such a structure. Since the SRRs can now be shifted without overlapping with a resonator from another unit cell, we can investigate the continuing trend in the response for $L_{shift}>0.75L_o$ (30µm). Figure 4(a) shows that the two resonances which appear at $L_{shift}=0.375L_o$ (15µm), trend to the uncoupled, "bare" resonance frequencies of the individual SRRs. Fig. 4(b) shows the magnitude of the on-resonance surface current densities in both SRRs for $L_{shift}=0$µm. Clearly, the two SRRs are responding to the incident electrical excitation as one resonator. The two SRRs are strongly coupled in this regime. However, when Fig. 4(b) is compared to Figs. 4(c) and 4(d), significantly different behavior is observed. Figs. 4(c) and 4(d) show the surface current densities in both SRRs for $L_{shift}=1.5L_o$ at f=0.8THz and f=1.1THz, respectively. It is apparent that the 0.8THz resonance is driven by currents excited predominately in the small gap SRR, while the 1.1THz resonance is driven by currents excited predominately in the large gap SRR. Thus, when $L_{shift}$ is increased beyond $0.375L_o$, the system trends to a state where the individual SRRs respond to incident radiation as separate elements. Thus, there are two separate resonant modes, with frequencies determined by the sub-geometry of the individual SRRs and not their relative lateral displacement. The SRRs are now effectively individual, decoupled resonators for $L_{shift}/L_o>0.375$.

Many aspects of the ABC-SRRs' two mode response become intuitive when the conceptual models of [17, 19] are applied. The capacitive coupling depends on the charge distribution in the individual SRR gaps and the SRRs lateral displacement [17]. As the gap in a ring is made larger, the charge buildup in the SRR gap will decrease in magnitude. This effectively lowers the capacitive coupling between the SRRs, making the SRRs more prone to decoupled behavior for larger $\Delta g$. This explains the $\Delta g$ dependence for the strength of the higher frequency resonance.

The onset point of the crossover can also be understood conceptually. As the two SRRs are shifted laterally, the magnetic and electric coupling parameters defined in ref. [19] both become zero for a shift value corresponding to $L_{shift}=0.375L_o$ in this work. It is at this point of minimum interaction where the ABC-SRRs are able to transition from coupled to decoupled behavior and the two mode state appears. As the two SRRs are shifted beyond $L_{shift}=0.375L_o$, the coupling parameters briefly increase before asymptotically approaching 0 as the ABC-SRRs are shifted infinitely far apart. This increase in coupling gives rise to a "transition region" between the one and two mode states. In this region, clearly visible in Fig. 3 and 4(a), two resonances appear but still shifted slightly from the bare SRR resonance frequencies.

Finally, while this decoupling transition was not observed in the previous studies on symmetric BC-SRRs [17, 19] the results presented in this letter are consistent with previous work. For the symmetric ($\Delta g=0\mu m$) BC-SRR structures considered previously, decoupled behavior is not observed for two reasons. First, the two uncoupled resonators are degenerate, allowing for only one electric mode at all shift values. Additionally, as the previous work focused on square unit cells with small periodicity, it was impossible to shift the BC-SRRs out of the transition region discussed above. The net effect is a resonant response with one mode that exhibits coupled behavior for all accessible values of $L_{shift}$.

In summary, we investigated the response of asymmetric BC-SRRs (ABC-SRRs) under lateral shift using terahertz time-domain spectroscopy and numerical simulations.

We observe a transition from a one resonance state to a two resonance state for shift values larger than $L_{shift}/L_o=0.375$ where $L_o$ is the side length of an SRR. For lateral shifts lower than this value, the ABC-SRRs act as one coupled resonant element. Above this value, the component SRRs respond to incident radiation as separate, uncoupled resonators, as evidenced by the simulated on-resonance surface current densities. While this behavior is unique to ABC-SRRs, it is consistent with previously published results for symmetric BC-SRR structures and can be explained using similar conceptual models. As substrate induced bianisotropy, fabrication error, and other effects can conspire to make symmetric BCSRR effectively asymmetric, these results provide a description of ABC-SRR behavior essential for complete understanding of BC-SRR based metamaterials.

The authors acknowledge support from DTRA C&B Technologies Directorate administered through a subcontract from ARL, and from NSF under contract number ECCS 0802036 and AFOSR under contract number FA9550-09-1-0708. The authors would also like to thank the Photonics Center at Boston University for technical support throughout this project.

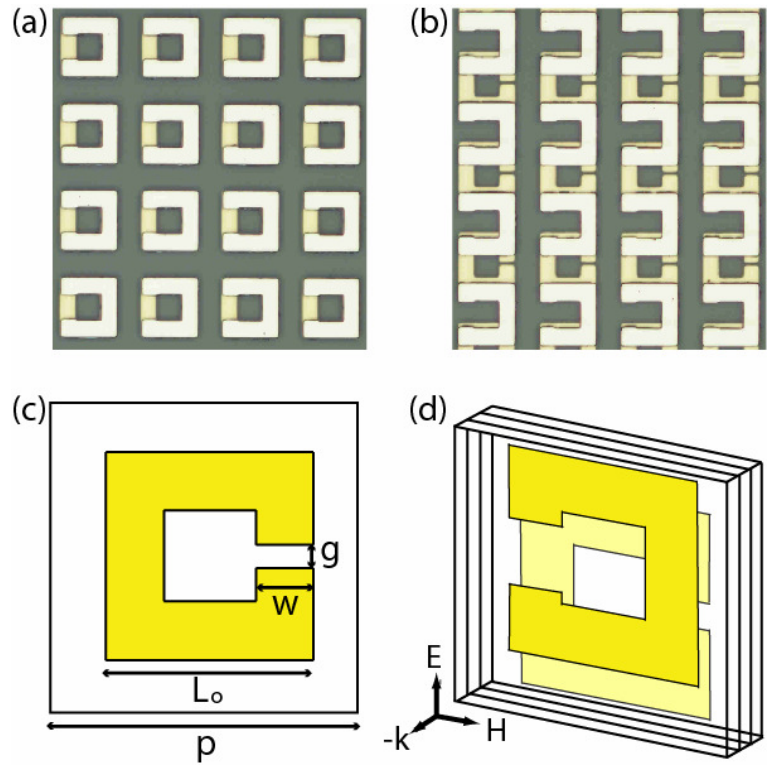

**Figure 1:** (**a**) Photograph of unshifted ABC-SRR array. The front gap is g=16 μm, while the back gap is $g_{back}$=2μm. (**b**) Photograph of shifted ABC-SRR array. The dimensions are the same as in (a). (**c**) Top-down schematic of ABC-SRR unit cell, showing the top ring and relevant dimensions. (**d**) Perspective view of ABC-SRR unit cell showing the direction of lateral shifting and the polarization and direction of the THz signal used to excite the metamaterials.

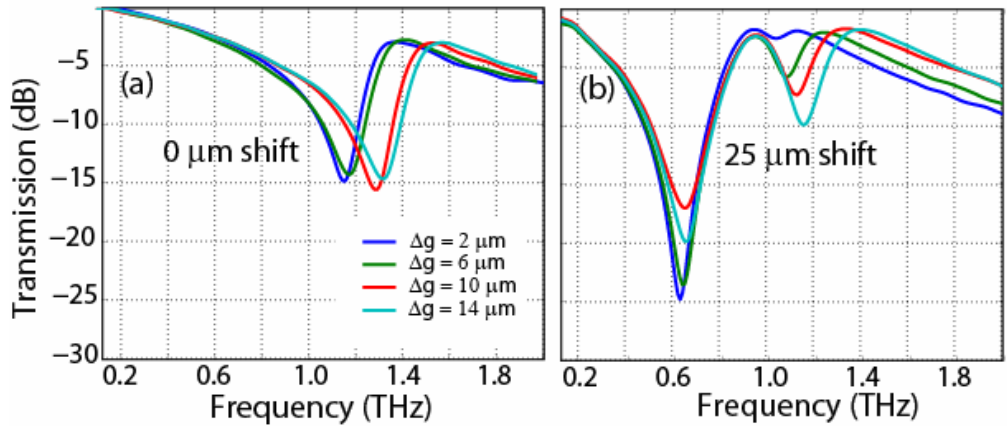

**Figure 2: (a)** Experimental transmission curves for 0µm shifted ABC-SRRs with varying gap differences. In this unshifted case, only the fundamental electrical resonance of the BC-SRR is excited. **(b)** Experimental transmission curves for the corresponding 25µm shifted ABC-SRRs. Here, the electrical resonance has redshifted, consistent with [17]. However, a new mode appears at higher frequencies and is strongly dependent on the asymmetry between the SRRs. Plots are scaled in dB for clarity, as the resonances are very wide when plotted on a linear scale.

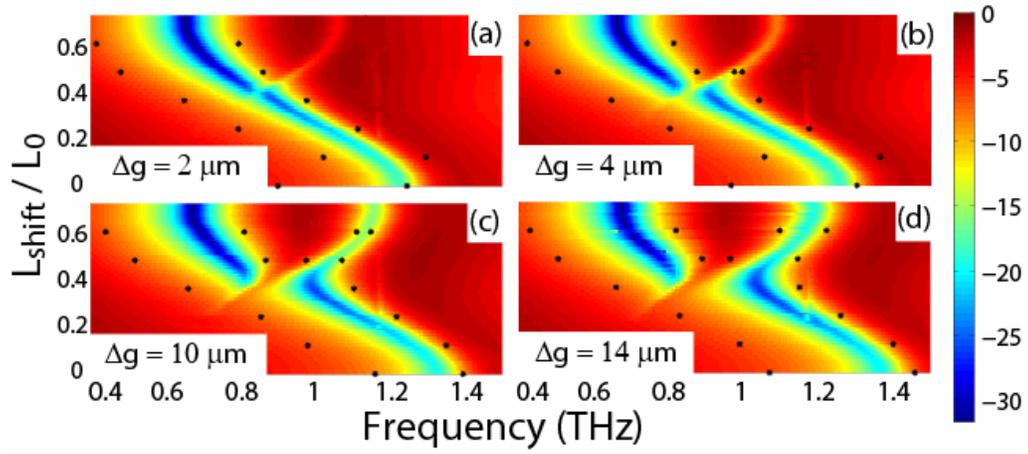

**Figure 3:** Simulation results showing Transmission (in dB) curves vs. shift for (**a**) 2µm, (**b**) 4µm, (**c**) 10µm, and (**d**) 14µm gap differences. All frequency points are normalized to the resonance of an isolated SRR with 2µm wide capacitive gap and other dimensions as given in Figure 1 c). Shift Distances are normalized to the side length of the SRRs, $L_o$. Asterisks denote the experimental -7dB points and are included to show correspondence of simulation with experiment.

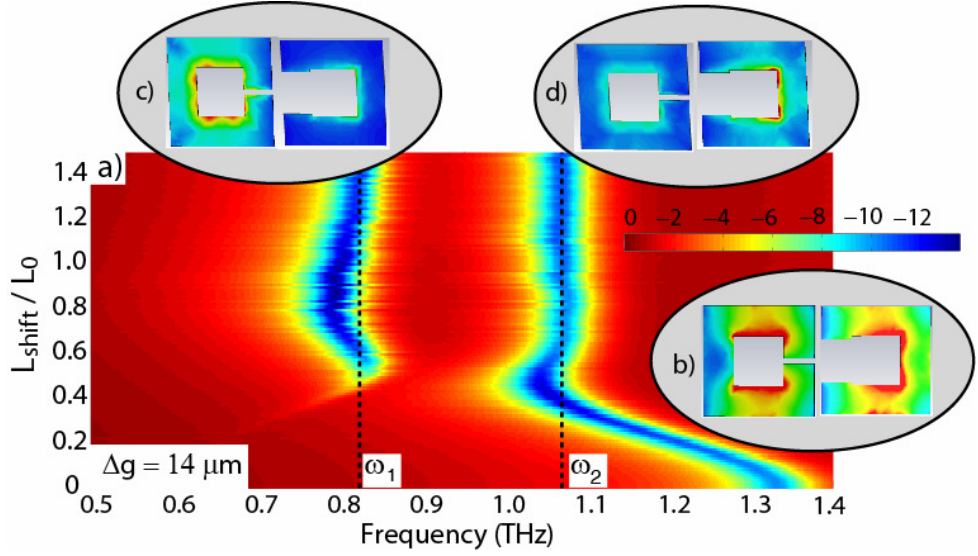

**Figure 4: a)** Simulation results showing transmission curves (in dB) vs. lateral shift for an ABC-SRR with a 14µm gap difference in a rectangular unit cell lattice ($P_y=2P_x=120\mu m$). All other dimensions are the same as in the previous structures. The dotted lines at $\omega_1$ and $\omega_2$ correspond to the uncoupled resonance frequencies of the smaller gap SRR and the larger gap SRR, respectively. Shift distances are normalized to the side length of the SRR, $L_o$. In this large unit cell lattice the SRRs can be shifted far enough apart to completely decouple from each other and from SRRs in the nearest neighbor cells. As the SRRs are shifted, the two resonances trend to the uncoupled resonance frequencies. **b)** On resonance surface current density in arbitrary units in both SRRs for $L_{shift}=0L_o$, showing that at the 1.3THz resonance the SRRs are equally excited, and acting as one coupled resonator. **c)** On resonance surface current density in arbitrary units in both SRRs for $L_{shift}=1.5L_o$ and f=0.8THz, showing that the response is dominated by currents in the small gap SRR. **d)** On resonance surface current density in arbitrary units in both SRRs for $L_{shift}=1.5L_o$ and f=1.1THz, showing that the response is dominated by currents in the large gap SRR. Thus, for large shifts, the SRRs are excited as separately, as uncoupled resonators.